\documentclass[hyper,a4paper,12pt]{JHEP3}
\usepackage{graphicx}
\usepackage{epsfig}
\usepackage{amsmath}
\usepackage{amssymb}
\usepackage{amsthm}

\usepackage{ulem}

\def\unit{{1\kern-.65ex {\rm l}}}
\def\1{{1\kern-.65ex {\rm l}}}

\let\vev=\bracket

\def\CL{{\cal L}}
\def\CM{{\cal M}}

\def\CO{{\cal O}}

\title{Holographic Brownian Motion in Two Dimensional Rotating Fluid}
\author{Ardian Nata Atmaja$^{1,2}$\\
${}^1$Research Center for Physics, Indonesian Institute of Sciences (LIPI)\\
~Kompleks PUSPITEK Serpong, Tangerang 15310, Indonesia.\\
${}^2$Indonesia Center for Theoretical and Mathematical Physics (ICTMP)\\
~Bandung 40132, Indonesia.\\
email: \email{ardian@teori.fisika.lipi.go.id}
}

\abstract{The Brownian motion of a heavy quark under a rotating plasma corresponds to BTZ black hole is studied using holographic method from string theory. The position of heavy quark is represented as the end of string at the boundary of BTZ black hole and the corresponding rotating plasma is one dimensional compact space. We requires the angular velocity of the string fluctuation to be equal to the ratio between inner horizon and outer horizon, called terminal angular velocity, which is related to the zero total force condition. We show the displacement square of this solution behaves as a Brownian particle in non-relativistic limit. For relativistic case, we argue that it is more appropriate to compute the leading order of low frequency limit of random-random force correlator. The Brownian motion relates this correlator with physical observables: effective mass of the Brownian particle, friction coefficient, and temperature of the plasma.}

\preprint{}
\begin{document}

\section{Introduction}

AdS/CFT correspondence has been widely used to solve some problems in non-perturbative gauge theories. Some of the applications of AdS/CFT correspondence is related to heavy ion collisions experiments at RHIC and LHC~\cite{CasalderreySolana:2011us}. The original proposal of AdS/CFT correpondence was formulated in the context of strongly coupled supersymmetric gauge theory and even more it was restricted to conformal field theory~\cite{Maldacena:1997re}. However, it is believed that at very high energy all gauge theories, at least with gravity dual, share some universalities~\cite{Janik:2010we}. One of the example is that the universality in the ratio between shear viscosity and entropy density~\cite{Kovtun:2004de}.

A Quark Gluon Plasma (QGP) is created when two heavy ions collide at very high temperature. In this new state of matter, hadrons of the two heavy ions are disolved into a soup of quarks and gluons. A relatively heavy particle, e.g. a heavy quark, moves randomly in QGP as it collides with the constituent of QGP, quarks and gluons. Its motion can be described by a well-known Brownian motion if its time evolution of the displacement square, $s(t)\equiv x(t)-x(0)$, is given as follows~\cite{Uhlenbeck:1930zz}
\begin{equation}
\label{displacement}
 \vev{s(t)^2}\approx \left\{
\begin{array}{ll}
   {T\over m}t^2 &~~~  (t\ll t_{\rm relax}) \\
  2Dt            &~~~  (t\gg t_{\rm relax}) 
\end{array}\right. ,
\end{equation}
where
\begin{equation}
 t_{\rm relax}\sim{1\over \gamma_0},~~~ \gamma_0\equiv\int_0^\infty dt ~\gamma(t),
\end{equation}
 is the relaxation time of the Brownian particle, $\gamma(t)$ is the memory kernel function~\cite{Mori:genLE},
\begin{equation}
 D= {T \over \gamma_0 m}
\end{equation}
is the diffusion constant, $T$ is the temperature, and $m$ is the mass of Brownian particle. At early time, $t\ll t_{\rm relax}$, the Brownian particle is in balistic regime where the particle moves with constant velocity. At the late time, $t\gg t_{\rm relax}$, the particle is in diffusive regime and undergoes a random walk after collides with many particles of plasma's constituent.

In particular the QGP that we would like to consider is strongly coupled and hence we will use AdS/CFT correspondence to compute the displacement square. In general, the AdS/CFT correspondence is a correspondence between strongly coupled gauge theory in $d$ dimension and weakly coupled gravity theory in $d+1$ dimension, with the background metric is Asymptotically AdS. An observable, $\CO$, of gauge theory, which is hard to be computed in strongly coupled system, can be easily computed in the corresponding weakly coupled gravity theory by solving equation of motion of the dual field, $\phi$, in the gravity theory where its boundary value, $\phi(r\to\infty)$,  acts as a source for the observable of gauge theory~\cite{Gubser:1998bc,Witten:1998qj},
\begin{equation}
 \vev{e^{\int_{\partial \CM}\CO \phi_0}}= e^{\int_{\CM}\CL^{\rm gravity}_{\rm on-shell}[\phi(r\to\infty)\equiv \phi_0]},
\end{equation}
with the boundary $\partial \CM$ where the gauge theory lives in is at $r\to\infty$. At the present, we still don't have the corresponding gravity theory of QCD, the microscopic theory of QGP. However, with universalities of these gauge theories as mentioned above, we might be able to get some features of QGP qualitatively by computing using available gravity theories. 

There are many calculations have been done with gravity theories where the corresponding plasma is not rotating. However, it is also natural to think that QGP should also posses some angular momentum, see~\cite{Becattini:2007sr} and reference there in. In this
paper we would like to study the Brownian motion of an external quark in two
dimensional rotating plasma where the corresponding gravity is three dimensional
gravity with the background metric is the BTZ black hole~\cite{Banados:1992wn}. There are two main methods in calculating the holographic Brownian motion. The first method is done by using the so-called Hartle-Hawking state in the Hawking radiation near the horizon and with interpretation that the end of string described position of the external quark~\cite{deBoer:2008gu}. The second method interprets the end of string as the source for total force in the fluid acting on the external quark, following the GKPW procedure~\cite{Gubser:1998bc,Witten:1998qj}, and the calculation uses some extension of the Scwhinger-Keldish formalism~\cite{Son:2009vu}. It is not clear how both methods can describe the same physical result on the Brownian motion at the boundary. The connection between these methods might give more understanding on the Hawking radiation process and it is worth for studying. More detail comparisons about these two methods can be found in~\cite{Hubeny:2010ry}. In this paper, we are going to use the first method to evaluate the asymptotic time behaviour of displacement square of the external quark in two dimensional rotating plasma dual to BTZ black hole background in the gravity theory\footnote{There has been study of the holographic Brownian motion in the presence of a magnetic field in the plasma and its dual non-commutative plasma~\cite{Fischler:2012ff}.}.

\section{Gravity Setup}

An external quark moving in a plasma can be presented by a fundamental string dangling from the boundary to the horizon of the background metric where the end of string at the boundary acts as a source for an external quark~\cite{Karch:2002sh}. The length of this string related to the mass of the external quark. Therefore if the boundary is located near infinity, $r\to\infty$, the external quark is very heavy. The dynamic of this string is classically given by the Nambu-Goto action
\begin{eqnarray}
 S_{NG}&=&-{1 \over 2\pi\alpha'}\int d\sigma^2 \sqrt{(\dot{X}. X')^2-\dot{X}^2 X'^2},
\end{eqnarray}
with
\begin{eqnarray}
 \dot{X}. X'&=&g_{\mu\nu}{dX^\mu \over d\sigma^0}{dX^\nu \over d\sigma^1},\\
\dot{X}^2&=&g_{\mu\nu}{dX^\mu \over d\sigma^0}{dX^\nu \over d\sigma^0},\\
X'^2&=&g_{\mu\nu}{dX^\mu \over d\sigma^1}{dX^\nu \over d\sigma^1},
\end{eqnarray}
where $\sigma\equiv(\sigma^0,\sigma^1)$ is the string worldsheet coordinate, $X^\mu$ is the spacetime coordinate, $\alpha'$ is related to the string length scale, and $g_{\mu\nu}$ is the background metric. 

The background metric, $g_{\mu\nu}$, is given by BTZ black hole~\cite{Banados:1992wn}:
\begin{align}
\label{BTZ metric}
 ds^2&=-N^2dt^2+\frac{1}{N^2}dr^2+r^2\left(N^\phi dt+d\phi\right)^2,\nonumber\\
N^2(r)&=-M+\frac{r^2}{l^2}+\frac{J^2}{4r^2},\ \ \ \ \ N^\phi (r)=-\frac{J}{2r^2},
\end{align}
with $-\infty<t<\infty,~0<r<\infty$ and $0\leq\phi< 2\pi$. $M$ and $J$ are the conserved charges with asymptotic invariance under time displacement (mass) and rotational (angular momentum) invariance. The Hawking temperature of BTZ black hole, which is interpreted as the temperature of the plasma, is given by
\begin{align}
\label{temperature}
 T&=\frac{r_+^2-r_-^2}{2\pi l^2 r_+},\nonumber\\
 r_{\pm}=\sqrt{\frac{Ml^2}{2}\pm\frac{1}{2}\xi},&\ \ \ \ \xi=\sqrt{M^2l^4-J^2l^2},
\end{align}
where $r_-$ is the inner horizon and $r_+$ is the outer horizon. From now on we will set the length square dimension $l^2=1$ for convenient.

\section{Small Fluctuation}
The Nambu-Goto action under the BTZ black hole, with axial gauge $(\sigma^0,\sigma^1)=(t,r)$, becomes
\begin{align}
 S_{NG}=&-\frac{1}{2\pi\alpha'}\int d\sigma^2 \left[\left(\dot{\phi}+N^\phi\right)^2r^4\phi'^2-\left(N^{-2}+r^2\phi'^2\right)\left(-N^2+r^2\left(\dot{\phi}+N^{\phi}\right)^2\right)\right]^{1/2}.
\end{align}
Under this axial gauge, we are left with one degree of freedom. So, the equation of motion for $\phi$ is the following:
\begin{align}
\label{EOM}
 -\frac{\partial}{\partial t}\left[\frac{r^2N^{-2}}{\sqrt{-g}}\left(N^\phi+\dot{\phi}\right)\right]+\frac{\partial}{\partial r}\left[\frac{r^2\phi'N^2}{\sqrt{-g}}\right]=0.
\end{align}

\subsection{Trivial Solution}
The trivial solution to the equation of motion (\ref{EOM}) is given by $\phi(\sigma)=C$ where $C$ is an arbitrary constant. If we take small fluctuation under this trivial solution such that $\phi=\phi+C$, $|\partial_t\phi|\ll 1$, and $|\partial_r\phi|\ll1$, the Nambu-Goto action up to quadratic terms is given by
\begin{align}
 S_{NG}\approx&-\frac{1}{4\pi\alpha'}\int d\sigma^2\frac{1}{\sqrt{N^2\left(r^2-M\right)}}\left(-\frac{N^2r^2}{\left(r^2-M\right)}\dot{\phi}^2+r^2N^4\phi'^2\right),
\end{align}
where we have kept only the quadratic terms that contribute to equation of motion. The equation of motion of the above Nambu-Goto action is
\begin{align}
\label{EOM trivial}
-\frac{N^2r^2}{\left(r^2-M\right)\sqrt{N^2\left(r^2-M\right)}}\ddot{\phi}+\frac{\partial}{\partial r}\left(r^2N^2\sqrt{\frac{N^2}{r^2-M}}\phi'\right)=0.
\end{align}
Writing the solution as $\phi(t,r)=e^{-i\omega t}f_\omega(r)$ and changing coordinates to $s=r^2-\frac{1}{2}M$, the equation of motion now becomes
\begin{align}
\partial^2_s f_\omega +\left(\frac{3}{2(s-Q)}+\frac{3}{2(s+Q)}-\frac{1}{2(s-P)}\right)\partial_s f_\omega +\frac{\omega^2}{4(s-Q)(s+Q)(s-P)}f_\omega=0,
\end{align}
with $P=\frac{1}{2}M$ and $Q^2=\frac{1}{4}\xi^2$. In general, the solution to this equation is very complicated. However for the extremal case, with $|J|=M$ or $Q=0$, the solution can be found analytically and it is given by
\begin{align}
\label{trivial sol}
 f_\omega^{(\pm)}=s^{-1\mp D}{}_2F_1(-1\mp D,\frac{1}{2}\mp D;1\mp 2D;\frac{s}{P}),
\end{align}  
with $D=\frac{1}{P}\sqrt{P\left(P+\frac{1}{4}\omega^2\right)}$ and ${}_2F_1$ is hypergeometric function. 

We may discard the solutions (\ref{trivial sol}) by two reasons.  First, the Brownian computation we use here based on Hawking radiation near the outer horizon and that requires the fluctuations to have oscillatory modes of the form $e^{-i\omega(t\pm r_*)}$, where $r_*\equiv r_*(r)$ is real everywhere and it is usually the tortoise coordinate. One can see that near the outer horizon, $s\to 0$, the hypergeometric function in solutions (\ref{trivial sol}) equals to one and the solutions behave like $e^{\left(-1\mp\sqrt{1+{\omega^2\over 4 r_+}}\right)\ln (r-r_+)}$. Hence the solutions (\ref{trivial sol}) can not have oscillatory modes in radial coordinate unless $\omega$ is complex but this makes the modes become dissipative in the lower half plane of frequency space, the modes die off as the time increases, which is characteristic of quasinormal modes and this is not what we expect from the motion of Brownian particle. In general, we can see this when rewriting the equation (\ref{EOM trivial}) into the form of Schr\"{o}dinger equation by a coordinate transformation to tortoise coordinate
\begin{align}
 dr_*={1\over r^2 N^2}\sqrt{r^2-M \over N^2}dr={1\over (r^2-r_-^2)(r^2-r_+^2)}\sqrt{r^2(r^2-r_+^2-r_-^2) \over (r^2-r_+^2)(r^2-r_-^2)}dr.
\end{align}
Near the outer horizon, $r\to r_+$, the tortoise coordinate is imaginer. However, we could resolve it by putting $i$ in the above transformation and we will find the near outer horizon solutions of (\ref{EOM trivial}) are given by $\phi= e^{-i\omega t} e^{\pm A \omega r_*}$, where $A$ is a positive number. For the same reason as before, $\omega$ needs to be complex in order to have oscillatory modes in spacetime. So, we may discard the fluctuation around this trivial constant solution because of the presence of these quasinormal modes and furthermore the tortoise coordinate above is ill-defined since it is not real everywhere. Second, the trivial constant solution itself is not physical because square root determinant of the worldsheet metric
\begin{align}
 \sqrt{-g}=\sqrt{1-{r_+^2 r_-^2 \over (r^2-r_+^2)(r^2-r_-^2)}}
\end{align}
is not real everywhere. As discussed in~\cite{Herzog:2006gh}, imaginary value of the above square root determinant could be a signal of superluminal propagation which leads to the violation of causality. It also causes the action, energy, and momentum to become complex hence we must discard this solution. So it is unreasonable in the first place to fluctuate around this trivial constant solution.

\subsection{Linear Solution}
\label{linear solution}
A non-trivial solution is given by the following linear ansatz
\begin{align}
 \phi(t,r)=w t+\eta(r),
\end{align}
where $w$ is a constant angular velocity. Substitute this into equation (\ref{EOM}) the solution for $\eta$ is
\begin{align}\label{linear sol}
 \eta'(r)=-\frac{\pi_\phi}{r^2N^2}\sqrt{\frac{r^4\left(w+N^\phi\right)^2-r^2N^2}{\pi_\phi^2-r^2N^2}},
\end{align}
where $\pi_\phi$ is a constant related to momentum conjugate of $\phi$ in $r$ direction and is also interpreted as the total force experienced by the quark~\cite{Herzog:2006gh,Gubser:2006bz}. The negative sign in (\ref{linear sol}) is chosen to describe a drag force and it is characterized by a single number, $\pi_\phi$, which is related to the total force applied to the string to keep it moving with linear angular velocity, $w$. That number is fixed by the requirement that the string solution (\ref{linear sol}) is real everywhere along the worldsheet. Since the numerator of the square root vanishes at some $r\equiv r_{NH}$, reality requires that the denominator also vanishes there, giving the condition described, $\pi^2_\phi= r^2_{NH} N(r_{NH})^2$ with $r^2_{NH} = {M-Jw \over 1-w^2}$ and a constraint $w^2<1$. One can show that $r^2_{NH}\geq r^2_+$ and, for $r_{NH}\neq r_+$, it does not correspond to the horizons of the induced metric on the worldsheet. 

This non-trivial solution also has static solution at $w=0$ but its not equal to the above trivial solution. One can check that this static solution is curved with
\begin{align}
 \eta'(r)=-{r_+^2 r_-^2 \over (r^2-r_+^2)(r^2-r_-^2)}
\end{align}
and
\begin{align}
 \pi_\phi^2=r_+^2 r_-^2={J^2 \over 4}.
\end{align}
Therefore the static string feels some external force caused by the rotation of the black hole as much as 
\begin{align}
F_{ext}={\pi_\phi\over 2\pi\alpha'}={J\over 4\pi\alpha'}.
\end{align}
We must extract this external force in order to get the friction coefficient for non-zero $w$ which is given by
\begin{align}
\label{rotation friction}
 \gamma_0m_0={r_+^2+r_-^2-2r_+r_-w \over 2\pi\alpha'(1-w^2)}={r_{NH}^2 \over 2\pi\alpha'}.
\end{align}
In deriving the friction coefficient, we have assumed that the momentum is non-relativistic, $P_\phi=m_0 w$, where $m_0$ is non-relativistic mass of the external quark.

The equation of motion from the Nambu-Goto action after substituting the small fluctuation under this background, $\phi\to w t+\eta(r)+\phi$, up to quadratic terms is quite complicated. One can check that for almost all values of $w$ the solution does not have oscillatory modes in time since the equation contains terms with first derivative in time. However, there are values of $w$ in which the equation of motion becomes simple and they are solutions of $J-2M w+J w^2=0$. We pick a solution $w = r_-/r_+$ since it is consistent with the constraint $w^2 < 1$. This value turns out to be equal to the velocity where the total force acts on the external quark, or the momentum flux on the string, vanishes in the drag force formula~\cite{NataAtmaja:2010hd}. For this value of angular velocity, which we shall call now as terminal angular velocity, the special radius $r_{NH}$ approaches the outer horizon of the BTZ black hole, where $N( r_+)=0\to \pi_\phi=0$, so that the steady-state solution is that with $r_{NH} = r_+$. The equation of motion for this terminal angular velocity is simply given by
\begin{align}
\label{EOM linear}
 -\frac{r^2N}{\left(N^2-r^2(w+N^\phi)^2\right)^{3/2}}\ddot{\phi}+\frac{\partial}{\partial r}\left[\frac{r^2N^3\phi'}{\left(N^2-r^2(w+N^\phi)^2\right)^{1/2}}\right]=0.
\end{align}
Changing the coordinate to $s=r^2-r_+^2$, we obtain
\begin{align}
PQ\phi''+\left(\frac{1}{2}(PQ'-QP')+PQ'\right)\phi'-\frac{1}{4}\ddot{\phi}=0,
\end{align}
with 
\begin{align}
 P=N^2-r^2(w+N^\phi)^2=\frac{4\xi r_-^2}{J^2}s,\ \ \ \ \ Q=r^2N^2=\frac{1}{}(s+\xi)s.
\end{align}

The solution is obtain by writing $\phi(t,s)=e^{-i\omega t}f_\omega(s)$ and so the equation reduce to
\begin{equation}
PQ f_\omega''+\left(\frac{1}{2}(PQ'-QP')+PQ'\right)f_\omega'+\frac{\omega^2}{4}f_\omega=0.
\end{equation}
Linearly independent solutions are given by
\begin{equation}
\label{lin sol}
 f^\pm_\omega(s)=s^{\pm\frac{i\chi}{\sqrt{\xi}}}{}_2F_1({\pm\frac{i\chi}{\sqrt{\xi}}},\frac{3}{2}\pm\frac{i\chi}{\sqrt{\xi}};1\pm\frac{2i\chi}{\sqrt{\xi}};-\frac{s}{\xi}),
\end{equation}
with $\chi^2=\frac{\omega^2J^2}{16\xi r_-^2}$. Recalling that $r_+>r_-$, one can easily check that $\chi$ is real. The explicit form of these oscillation modes are
\begin{align}
 f^{\pm}_\omega=\frac{2^{\pm\frac{2i\chi}{\sqrt{\xi}}}}{(1\pm\frac{2i\chi}{\sqrt{\xi}})}\frac{\left(1\pm\frac{2i\chi}{\sqrt{s+\xi}}\right)}{\left(1+\sqrt{1+\frac{s}{\xi}}\right)^{\pm2\frac{i\chi}{\sqrt{\xi}}}}s^{\pm\frac{i\chi}{\sqrt{\xi}}}.
\end{align}
Asymptotic behaviour of the solutions near the outer horizon ($s\to 0$) and boundary ($s\to\infty$) is
\begin{align}
\label{asymptotic}
 f^\pm_\omega(s)\sim\left\{\begin{array}{ll}
e^{\pm i\omega s_*}, & (s\to 0)\\
\frac{(4\xi)^{\frac{\pm i\chi}{\sqrt{\xi}}}}{(1+\frac{\pm 2i\chi}{\sqrt{\xi}})}\left[1+\frac{2\chi^2}{s}+\cdots\right], & (s\to\infty),                                            \end{array}\right.
\end{align}
with $s_*=\frac{\chi}{\omega\sqrt{\xi}}\ln(s)=\frac{J}{4\xi r_-}\ln(s)$.

\section{Displacement Square}
We proceed by imposing boundary conditions to our solutions following procedure in~\cite{deBoer:2008gu}. We put an UV-cutoff near the boundary in order to have a finite large mass on the external quark such that it exhibits a Brownian motion. The other boundary condition, an IR-cutoff near the outer horizon, is to regulate the theory. Writing the solution
\begin{equation}
 f_\omega(s)=A\left[f^+_\omega(s)+B f^-_\omega(s)\right],
\end{equation}
with $A$ and $B$ are constants, we apply the Neumann boundary condition near the boundary, $\partial_sf_\omega(s_c)=0$ with $s=s_c\gg 0$, to implement the UV-cutoff and obtain
\begin{align}
 B&=\frac{(1-\frac{2i\chi}{\sqrt{\xi}})}{(1+\frac{2i\chi}{\sqrt{\xi}})}\frac{\xi+2i\chi\sqrt{s_c+\xi}}{\xi-2i\chi\sqrt{s_c+\xi}}\frac{(4s_c)^{\frac{2i\chi}{\sqrt{\xi}}}}{\left(1+\sqrt{\frac{s_c+\xi}{\xi}}\right)^{\frac{4i\chi}{\sqrt{\xi}}}}\equiv e^{i\theta_\omega}.
\end{align}
Hence the constant $B$ is merely a phase.
Imposing the same Neumann boundary condition near the outer horizon at $s_h=\epsilon$, $\epsilon\ll1$, we obtain
\begin{equation}
 B\approx \epsilon^{\frac{2i\chi}{\sqrt{\xi}}}=e^{-\frac{2i\chi}{\sqrt{\xi}}\ln(1/\epsilon)}.
\end{equation}
Imposing the second boundary condition will make the value of $\chi$ discrete and it is given by $\Delta\chi=\frac{\pi\sqrt{\xi}}{\ln(1/\epsilon)}$. In terms of $\omega$, the discreteness is
\begin{equation}
\label{descreteness}
 \Delta\omega=\frac{4\pi\xi r_-}{J}\frac{1}{\ln(1/\epsilon)}.
\end{equation}

\subsection{Hawking Radiation}
In this section we are going to use the formula of Hawking Radiation of the string transverse modes near the outer horizon of BTZ black hole to describe the random motion of the external quark~\cite{Lawrence:1993sg,Rey:1998ik,Myers:2007we}. As explained before, we are only interested in particular value of $w=r_-/r_+$ such that the Nambu-Goto action becomes
\begin{align}
\label{NG right omega}
 S^{(2)}_{NG}=-\frac{1}{4\pi\alpha'}\int dtdr\frac{r^2N^3\phi'^2}{\left(N^2-r^2(w+N^\phi)^2\right)^{1/2}}-\frac{r^2N\dot{\phi}^2}{\left(N^2-r^2(w+N^\phi)^2\right)^{3/2}}.
\end{align}
Write the general solution as
\begin{align}
\label{general sol}
 \phi(t,r)&=\sum^\infty_{\omega=-\infty}e^{-i\omega t}f_\omega(r),\\
f_\omega(r)&=g_\omega\left[f^+_\omega(r)+e^{i\theta_\omega}f^-_\omega(r)\right],
\end{align}
where the discreteness of $\omega$ is given in ($\ref{descreteness}$) and $g_\omega$ is a function of $\omega$ satisfying $g_{-\omega}=g^*_\omega$.
 
From (\ref{asymptotic}), near horizon ($s\sim 0$), we have
\begin{align}
\label{nh sol1}
 \phi(t,s\to 0)\sim\sum^\infty_{\omega=-\infty}g_\omega\left[e^{-i\omega(t-s_*)}+e^{i\theta_\omega}e^{-i\omega(t+s_*)}\right].
\end{align}
The right hand side in the above equation is divided into two terms. The first term is the ``UP'' modes and the second term is the ``IN'' modes. Because of Hawking radiation, the ``UP'' modes are excited. To see how much the ``UP'' modes are excited, we need to look at the Nambu-Goto action near the outer horizon. The Nambu-Goto action (\ref{NG right omega}) becomes, for $s\to 0$,
\begin{align}
 S^{(2)}_{NG}\sim\frac{J^2}{16r_-^2\pi\alpha'}\int dtds_*\left[\left(\frac{\partial\phi}{\partial t}\right)^2-\left(\frac{\partial\phi}{\partial s_*}\right)^2\right].
\end{align}
Redefining the field to $\Phi\equiv\frac{J}{2^{3/2}r_-\sqrt{\pi\alpha'}}\phi$ then the near horizon action becomes normalized free scalar field
\begin{equation}
 S^{(2)}_{NG}\sim\frac{1}{2}\int dtds_*\left(\dot{\Phi}^2-{\Phi'}^2\right).
\end{equation}
For a canonically normalized free scalar field $\Phi$, the mode expansions are
\begin{align}
 S=&\frac{1}{2}\int dt\int_0^L dx\left(\dot{\Phi}^2-{\Phi'}^2\right),\nonumber\\
\Phi(t,x)=&\frac{1}{\sqrt{L}}\sum_{k=-\infty}^{\infty}\frac{1}{\sqrt{2\upsilon_k}}\left[a_k e^{i(-\upsilon_kt+kx)}+ a_k^\dagger e^{i(\upsilon_kt-kx)}\right],\nonumber\\
k=\frac{2\pi n}{L},&~~n\in\mathbb{Z},~~\upsilon_k=|k|,~~\left[a_k,a^\dagger_l\right]=\delta_{kl},\nonumber\\
H=&\sum_k \upsilon_k \left(a^\dagger_k a_k+\frac{1}{2}\right),
\end{align}
where $L$ is the length of the system and we are taking periodic boundary condition. Following procedure in~\cite{deBoer:2008gu}, we find $k$ is discretized as
\begin{equation}
 \Delta k=\frac{4\pi\xi r_-}{J}\frac{1}{\ln(1/\epsilon)}.
\end{equation}
Comparing this with $\Delta k=2\pi/L$, we also obtain the effective size of the system $L$ which is
\begin{equation}
 L=\frac{J}{2\xi r_-}\ln(1/\epsilon).
\end{equation}
Therefore, the near outer horizon solution is written as
\begin{align}
\label{nh sol2}
  \phi(t,s\to 0)&=\frac{2^{3/2}r_-^{3/2}}{J^{3/2}}\sqrt{\frac{2\xi \pi\alpha'}{\ln(1/\epsilon)}}\sum_k\frac{1}{\sqrt{2\upsilon_k}}&\left[a_k e^{i(-\upsilon_kt+ks_*)} + a^\dagger_k e^{i(\upsilon_kt-ks_*)}\right].
\end{align}
There are two modes in the solution: ``UP'' and ``IN'' modes. The ``UP'' modes are the $k>0$ modes and the expectation value of the occupation number is given by the Bose-Einstein distribution:
\begin{align}
\label{B-E distribution}
\langle a^\dagger_k a_l\rangle=\frac{\delta_{kl}}{e^{\beta\upsilon_k}-1},~~~~~k>0, 
\end{align}
with $\beta=\frac{2\pi r_+}{r_+^2-r_-^2}$.
The ``IN'' modes ($k<0$) just fall into the black hole where the the expectation value is zero.

Comparing (\ref{nh sol1}) and (\ref{nh sol2}), we can write the general solution near outer horizon as $k$ sum~\cite{deBoer:2008gu}
\begin{align}
  \phi(t,s\to 0)\sim\sum^\infty_{k=-\infty}\left[G_k e^{i(-\upsilon_kt+ks_*)} + G_k^* e^{i(\upsilon_kt-ks_*)}\right],
\end{align}
where
\begin{equation}
 G_{k>0}=g_k,~~~G_{k<0}=g^*_k e^{-i\theta_k},
\end{equation}
with
\begin{align}
\label{operator relation}
 g_k=\left\{\begin{array}{ll}
\frac{2^{3/2}r_-^{3/2}}{J^{3/2}}\sqrt{\frac{\xi \pi\alpha'}{\upsilon_k\ln(1/\epsilon)}}a_k & (k>0)\\
\frac{2^{3/2}r_-^{3/2}}{J^{3/2}}\sqrt{\frac{\xi \pi\alpha'}{\upsilon_k\ln(1/\epsilon)}}e^{-i\theta_k}a^\dagger_k & (k<0).
\end{array}\right.
\end{align}
Because $g^*_k=g_{-k}$, this means that $a_k$ and $a^\dagger_k=a_{-k}$ are related.

\subsection{Position of the Brownian particle}

The end point of $\phi$ coordinate at $s=R$ is given by
\begin{align}
 \phi_R\equiv\phi(t,s=R)=\sum_{\upsilon=-\infty}^\infty \frac{g_\omega e^{-i\omega t} 2^{1+\frac{2i\chi}{\sqrt{\xi}}}\left(1-\frac{2i\chi}{\sqrt{\xi}}\right)\xi R^{\frac{i\chi}{\sqrt{\xi}}}}{\left(1+\sqrt{1+\frac{R}{\xi}}\right)^{\frac{2i\chi}{\sqrt{\xi}}}\left(\xi-2i\chi\sqrt{R+\xi}\right)}.
\end{align}
Using the relation (\ref{operator relation}) and (\ref{B-E distribution}), we get
\begin{equation}
 \langle g^\dagger_\omega g_{\omega'}\rangle=\delta_{\omega\omega'}\langle a^\dagger_\omega a_{\omega'}\rangle\frac{8r_-^{3}\xi\pi\alpha'}{J^{3}|\omega|\ln(1/\epsilon)}. 
\end{equation}
Then, we can compute
\begin{align}
 \langle\phi_R(t)\phi_R(0)\rangle&=\sum_{\omega>0}\frac{32r_-^{3}\xi\pi\alpha'(\xi^2+4\chi^2\xi)}{J^{3}\omega(\xi^2+4\chi^2(R+\xi))}{1\over \ln(1/\epsilon)}\left(\frac{e^{i\omega t}}{e^{\beta\omega}-1}+\frac{e^{-i\omega t}}{1-e^{-\beta\omega}}\right)\nonumber\\
&\approx\frac{8r_-^2\alpha'}{J^2}\int_0^\infty \frac{d\omega}{\omega}\frac{\left(\xi^2+\frac{J^2}{4r_-^2}\omega^2\right)}{\left(\xi^2+\frac{J^2}{4r_-^2}\frac{R+\xi}{\xi}\omega^2\right)}\left(\frac{e^{i\omega t}}{e^{\beta\omega}-1}+\frac{e^{-i\omega t}}{1-e^{-\beta\omega}}\right).
\end{align}
The integral above is devergence. We need to regularize it by normal ordering the $a,a^\dagger$ oscillator ($:a_\omega a^\dagger_\omega:\equiv:a^\dagger_\omega a_\omega:$). The normal order correlator becomes
\begin{align}
 \langle:&\phi_R(t)\phi_R(0):\rangle=\frac{8r_-^2\alpha'}{J^2}\int_0^\infty \frac{d\omega}{\omega}\frac{\left(\xi^2+\frac{J^2}{4r_-^2}\omega^2\right)}{\left(\xi^2+\frac{J^2}{4r_-^2}\frac{R+\xi}{\xi}\omega^2\right)}\left(\frac{2\cos(\omega t)}{e^{\beta\omega}-1}\right).
\end{align}
The displacement squared, $s(t)^2\equiv\langle [\phi_R(t)-\phi_R(0)]^2\rangle$, is
\begin{align}
 s_{reg}(t)^2=&\frac{16r_-^2\alpha'}{J^2}\left[\frac{R}{R+\xi}I_1+\frac{\xi}{R+\xi}I_2\right],
\end{align}
with
\begin{align}
 I_1=4\int_0^\infty\frac{dx}{x(1+a^2x^2)}\frac{\sin^2(\frac{kx}{2})}{e^x-1},\ \ \ \ \ I_2=4\int_0^\infty\frac{dx}{x}\frac{\sin^2(\frac{kx}{2})}{e^x-1},
\end{align}
and we have defined
\begin{align}
 x=\beta\omega,\ \ \ \ \ k=\frac{t}{\beta},\ \ \ \ \ a^2=\frac{J^2(R+\xi)}{4r_-^2\xi^3\beta^2}.
\end{align}
The evaluation of these integrals  and their behavior for $R\gg 1$ or $a\gg 1$ can be found in Appendix B of~\cite{deBoer:2008gu}. Therefore, if $R\gg 1$, $s_{reg}(t)^2$ has the following behavior:
\begin{align}\label{dq}
 s_{reg}(t)^2=\left\{\begin{array}{ll}
 \frac{16r_-^3\xi^{3/2}\pi\alpha'}{J^3(R+\xi)^{1/2}\beta} t^2 +O\left(\frac{1}{R+\xi}\right) & (t\ll\beta)\\
 \frac{16\pi r_-^2\alpha'}{J^2\beta}t+ O\left(\mbox{log}\frac{t}{\beta}\right)& (t\gg\beta).
 \end{array}\right.
\end{align}

One can check that the displacement square (\ref{dq}) is consistent with the non-rotating case computed in~\cite{deBoer:2008gu} simply by setting $J=0$ or $r_-=0$, which also implies $w=0$. Using this we can write\footnote{Here, we write back $l^2$ by examining the dimension of all variables.}
\begin{eqnarray}
 s_{reg}(t)^2=\left\{\begin{array}{ll}
 \frac{\alpha'}{l^4\rho_c} t^2 +O\left(\frac{1}{\rho_c^2}\right) & (t\ll\beta),\\
 \frac{\alpha'\beta}{\pi l^4}t+ O\left(\mbox{log}\frac{t}{\beta}\right)& (t\gg\beta),
 \end{array}\right.
\end{eqnarray}
with $\rho_c^2=1+\frac{R}{Ml^2}$. This is similar to the one calculated in~\cite{deBoer:2008gu} by recalling that we are using $\phi$ coordinate which is dimensionless.

Next, we would like to write the displacement square (\ref{dq}) in terms of physical variabels. To do so, we compute the mass of external particle using the total energy and momentum of the string under the background metric (\ref{BTZ metric}) which are given by~\cite{Herzog:2006gh}
\begin{align}
 E=\frac{1}{2\pi\alpha'}\int dr~\pi^0_t,\ \ \ \ \ \ p_\phi=-\frac{1}{2\pi\alpha'}\int dr~\pi^0_\phi,
\end{align}
with
\begin{align}
 \pi^0_t=&\frac{(\phi')^2}{\sqrt{-g}}\left(g_{t\phi}^2-g_{tt}g_{\phi\phi}\right)-\frac{g_{rr}}{\sqrt{-g}}\left(g_{tt}+g_{t\phi}\dot{\phi}\right),\\
\pi^0_\phi =& -\frac{g_{rr}}{\sqrt{-g}}\left(g_{t\phi}+g_{\phi\phi}\dot{\phi}\right).
\end{align}
For linear solution, $\phi(t,r)=w t+\eta(r)$, with $w =r_-/r_+$ and so $\eta'=0$, we obtain
\begin{align}
 E=\frac{r_+}{2\pi\alpha'\xi^{1/2}}\left(\sqrt{R+\xi}-\sqrt{\xi}\right),\ \ \ \ \ p_\phi=\frac{r_-}{2\pi\alpha'\xi^{1/2}}\left(\sqrt{R+\xi}-\sqrt{\xi}\right).
\end{align}
The mass is then defined as
\begin{align}
 m_0^2=E^2-p_\phi^2={1\over (2\pi\alpha')^2}\left(\sqrt{R+\xi}-\sqrt{\xi}\right)^2.
\end{align}
Writing in terms of physical mass $m_0\approx{\sqrt{R+\xi}\over 2\pi\alpha'}$, for large $R$, and temperature $T$, asymptotic time behaviour of the displacement squared is
\begin{eqnarray}
\label{dq1}
 \hat{s}_{reg}(t)^2\approx\left\{\begin{array}{ll}
\left(\frac{4\xi r_-^2}{J^2}\right)^{3/2}\frac{T}{m_0} t^2 & (t\ll t_c),\\
 2Dt & (t\gg t_c),
 \end{array}\right.
\end{eqnarray}
where the diffusion constant is given by
\begin{equation}
 D=\frac{2\pi\alpha'}{r_+^2}T.
\end{equation}
So, if the Sutherland-Einstein relation holds then the relaxation time of Brownian particle is given by
\begin{equation}
 t_c={1\over\gamma_0}=\frac{2m_0\pi\alpha'}{r_+^2}.
\end{equation}
One can check that the friction coefficient is the same as in (\ref{rotation friction}) for terminal angular velocity. Hence we have shown that small fluctuation around the linear solution at the boundary of BTZ black hole, for terminal angular velocity, behaves as a Brownian particle although the result (\ref{dq1}), in the balistic regime, is different by a factor of $\left(\frac{4\xi r_-^2}{J^2}\right)^{3/2}$ compared to the standard formula of (\ref{displacement}).

\section{Random-Random Force Correlator}

The derivation of the displacement square in the previous section is rather tedious since we have to perform the integral over the whole frequency $\omega$. This is even more complicated if we consider the charge rotating black holes or higher dimensional black holes with more than one spatial coordinates for the Brownian motion. Another way to prove the Brownian motion of a particle which is by computing low frequency limit of the correlation function of random-random force in the Langevin equation. 

Asymptotic behavior of the displacement formula for Brownian motion can be derived from the Langevin equation\footnote{This standard Langevin equation can be obtained from the generalized Langevin equation by taking zero frequency limit of the memory kernel $\gamma(t)$~\cite{deBoer:2008gu}. Although it is constant, in general, the friction coefficient can depend on momentum or velocity as shown in~\cite{Herzog:2006gh} for the boundary metric where spacetime dimension $d\neq 4$.}
\begin{align}
\label{BM}
 {dp \over dt}=-\gamma_0 p+ R(t)+K(t),
\end{align}
where $\gamma_0$ is the friction coefficient, $R(t)$ is the random force, and $K(t)$ is the external force. The random force is the fluctuation part of the Langevin equation which takes the following properties~\cite{Uhlenbeck:1930zz}:
\begin{align}
 \vev{R(t)}=0,\ \ \ \ \vev{R(t)R(t')}=\rho(t-t')
\end{align}
or in frequency modes
\begin{align}
\vev{R(\omega)R(\omega')}=2\pi \delta(\omega+\omega') \tilde{\rho}(\omega).
\end{align}
Assuming that $\rho(t)$ is a very sharp maximum function at $t=0$ and the equipartition of energy, one will obtain~\cite{Uhlenbeck:1930zz}
\begin{align}
 \int_{-\infty}^{+\infty} \rho(t) dt=\lim_{\omega\to 0}\tilde{\rho}(\omega)= 2\gamma_0 m T,\label{random 2 general}
\end{align}
where $m$ is the mass of Brownian particle and $T$ is the temperature of the fluid. A detail derivation can be read in Appendix~\ref{Appendix 1} and we also show there that the Brownian displacement formula can be derived simply using the Langevin equation (\ref{BM}) and low frequency limit random-random force correlator (\ref{random 2 general}). So, to proof that our system behaves as Brownian particle we just need to compute the friction coefficient $\gamma_0$, which in holographic can be extracted from the drag force formula~\cite{Gubser:2006bz,Herzog:2006gh} or by turning on world-volume electric field~\cite{deBoer:2008gu}, and the mass of Brownian particle then compute $\tilde{\rho}(0)$ and check if the relation (\ref{random 2 general}) is satisfied.

\subsection{Non-rotating BTZ black hole}

As an example, we consider a non-rotating BTZ black hole with the following metric
\begin{align}
 ds^2&= - (r^2-M^2)dt^2+{1\over r^2-M^2} dr^2+r^2 dX^2,\\
T&={M\over 2\pi},
\end{align}
where $M$ the horizon and $T$ is the Hawking temperature. Taking fluctuation over the static solution, $\dot{X}=X'=0$, as such the collerator of two random force in frequency mode is given by~\cite{deBoer:2008gu}
\begin{align}
 \vev{R(\omega)R(\omega')}=2\pi\delta(\omega+\omega')\left[{1\over \alpha'}4\pi T^3+\CO(\omega)\right].
\end{align}
Therefore we obtain that
\begin{align}
 \tilde{\rho}(0)={1\over \alpha'}4\pi T^3.\label{random 2 nonrot BTZ}
\end{align}
From the drag force computation in~\cite{Herzog:2006gh}, we have in non-relativistic limit
\begin{align}
 \gamma_0 m_0={M^2\over 2\pi \alpha'.}
\end{align}
Substituting this to (\ref{random 2 nonrot BTZ}) and using the Hawking temperature formula, we get the same equation as in (\ref{random 2 general}).

\subsection{Rotating BTZ black hole}
As we mentioned before, the fluctuation is taken over terminal angular velocity related to the zero total force configuration. This is to simplify our computation as such only the fluctuation parts is present in the Langevin equation. Explicitly one can consider the position of the external quark into two parts: classical and fluctuation part, $X=X_{class}+\delta X$, which imply to separation of the momentum into $p=p_{class}+\delta p$. Hence the Langevin equation can be written as
\begin{align}
 {dp \over dt}=-\gamma_0~p_{class}+K(t)-\gamma_0~\delta p+R(t).
\end{align}
In the rotating case $K(t)\neq 0$ since the fluid is rotated due to the rotation of the black hole. To cancel $K(t)$, we must fixed the angular velocity such that $\gamma_0~p_{class}=K(t)$, or corresponds to $\pi_\phi=0$, which is given by the terminal angular velocity $w={r_-/ r_+}$ and it turns out that $r^2_{NH}\to r^2_+$ for this value of $w$, see section \ref{linear solution}.

The action (\ref{NG right omega}), or in particular the equation of motion (\ref{EOM linear}), can be derived by considering small fluctuation over a static classical solution with the following metric, which we shall call the ''effective`` metric,
\begin{align}
 ds^2=&-\sqrt{N^2(N^2-r^2(w+N^\phi)^2)}dt^2+{1\over \sqrt{N^2(N^2-r^2(w+N^\phi)^2)}} dr^2\nonumber\\
     &+{r^2N^2 \over N^2-r^2(w+N^\phi)^2}d\phi^2,
\end{align}
or in terms of $r_+$ and $r_-$ is written as
\begin{align}
\label{eff geom}
 ds^2=&-(r^2-r_+^2)\sqrt{(r^2-r_-^2)(r_+^2-r_-^2)\over r^2 r_+^2}dt^2+{1\over(r^2-r_+^2)}\sqrt{r^2 r_+^2\over (r^2-r_-^2)(r_+^2-r_-^2)}dr^2+\nonumber\\
&+{(r^2-r^2_-)r_+^2 \over (r_+^2-r_-^2)}d\phi^2.
\end{align}
It is easy to check that the Hawking temperature of this ''effective`` metric is equal to the Hawking temperature of rotating BTZ black hole. Although the ''effective`` metric (\ref{eff geom}) is static, it behaves a bit different from the non-rotating BTZ black hole at $r\to\infty$ and so it is not AdS at the boundary. We certainly have to check if this metric reproduces the same action to all order in the expansion of the Nambu-Goto action. However, in the present case we only need up to the second order expansion since our calculation is based on the solution of the equation of motion (\ref{EOM linear}) and the Hawking radiation near the horizon, which means the ''effective`` metric must reproduce the same Hawking temperature (\ref{temperature}). Moreover the holographic method in~\cite{deBoer:2008gu} used here in principal does not necessary require the boundary metric to be AdS\footnote{As an example, the holographic method in~\cite{deBoer:2008gu} has been used for computing the late-time dynamics of the probe string under the background metric where its boundary has either Schr\"{o}dinger or Liftshitz symmetries~\cite{Tong:2012nf,Edalati:2012tc}.}.

Now, let us check if the ''effective`` metric gives a same friction coefficient as in (\ref{rotation friction}) with $w=r_-/r_+$. A simple calculation can be done using the usual drag force formula on the ''effective`` metric background with a linear curved ansatz to extract the friction coefficient~\cite{Herzog:2006gh,Gubser:2006bz}. One can check that they are indeed the same. We also need to check if the friction coefficient computed using the holographic Brownian motion formula is also the same. In the holographic Brownian motion, we calculate the admittance as in~\cite{deBoer:2008gu}  by introducing electrical boundary term at the UV-cutoff, with the boundary condition at $\rho=\rho_c$ is given by
\begin{align}
\rho^4 \partial_\rho \phi={2\pi\alpha'\over r_+^3}{C\over D}F_{\phi t},                                                                                                                                                                                                                                                                                                                                                                                                                                                                                                                                                                                                                                                                                                                                                                                                                                                                                                                                                                                                                                                                                                                                                                                                                                                                                                                                                                                                                                                                                                                                                                                                                                                                                                                                                                                                                                                                                                                                                                                                                                                                                                                                                                                                                                                                                                                                                                                                                                                                    
\end{align}
where $C$ and $D$ are constants coming from the near boundary behaviour of the ''effective`` metric components, $g_{\phi\phi}\sim D r^2$ and $\sqrt{g_{tt}/g_{rr}}\sim r^2/C^2$, and the external force is taken to be $F_{\phi t}=E_0 e^{-i\omega t}$, with $E_0$ is constant. Using the matching technique~\cite{Harmark:2007jy} and recalling the outgoing modes are radiated randomly, one can check that these constants are canceled out at the leading order of the low frequency limit of the admittance such that the friction coefficient
\begin{align}
 \gamma_0=\lim_{\omega\to 0}{1\over\mu(\omega)}={r_+^2 \over 2m_0\pi\alpha'}
\end{align}
is equal to the friction coefficient (\ref{rotation friction}) at the terminal angular velocity, $w=r_-/r_+$. 

A formula for two-point function of random force collerators was derived in~\cite{Atmaja:2010uu} for a general static metric
\begin{align}
 ds^2=-h_t(r) f(r) dt^2+{h_r(r)\over f(r)} dr^2+ G(r) dX^2,                                                                                                                   
\end{align}
where $h_r,h_t,G$ are all finite and non vanishing in $r_+\leq r<\infty$, and $r_+$ is the largest positive root of $f(r)$. At low frequency limit, the random-random force collerator can be written as
\begin{align}
 \langle R(\omega) R(\omega')\rangle=2\pi \delta(\omega+\omega')\left[{G(r_+) T_H \over \pi\alpha'}+\mathcal{O}(\omega)\right],
\end{align}
where $T_H$ is the Hawking temperature of the general static metric. Hence for the ''effective`` metric (\ref{eff geom}) we can compute\footnote{Unlike in~\cite{Atmaja:2010uu}, the ''effective`` metric has $G\sim D r^2$ and $f\sqrt{h_t\over h_r}\sim r^2/C$ near the boundary. However, similar to the impedance the constants $C$ and $D$ do not appear at the leading order of the low frequency limit of random-random force correlator.} 
\begin{align}
 \tilde{\rho}(0)={r_+^2T\over \pi\alpha'}=2\gamma_0 m_0 T,
\end{align}
and once more it satisfies the formula (\ref{random 2 general}) which implies the Brownian motion of
the external quark.

\section{Discussion and Conclusion}
The displacement square of string fluctuation at the boundary, (\ref{dq1}), asymptotically showed behaviours like a Brownian particle. In comparison with the standard asymptotic behaviour, its early times gets multiplied by a factor of
\begin{align}
\left(\frac{4\xi r_-^2}{J^2}\right)^{3/2}.
\end{align}
We can write this factor in terms of terminal angular velocity $w=r_-/r_+$ such that the multiplication factor becomes $(1-w^2)^{3/2}$. Notice that the Brownian motion formula was originally derived under non-relativistic assumption, see Appendix \ref{Appendix 1}. Therefore in non-relativistic limit, $w\to0$, we get back the displacement square as in (\ref{displacement})\footnote{Note that since the boundary metric is two-dimensional ``Minkowski'', our Holographic calculation by itself is relativistic.}.

An interesting question is whether the relativistic version of the displacement square still have the same formula as in (\ref{displacement}) or it gets modified by some power of Lorentz factor, $\gamma\equiv1/\sqrt{1-w^2}$. In relativistic, the friction coefficient computed from the drag force formula is multiplied by a factor of $\gamma^{-1}$ where the momentum is defined as $P_\phi\equiv m_r w=\gamma m_0 w$. However, the quantity $\gamma_0 m_0$ does not change its form in the relativistic version since $\gamma_0\sim m_0^{-1}$ and so the diffusion constant is given by the same formula
\begin{align}
 D={T\over \gamma_r m_r},~~~~\gamma_0 m_0=\gamma_r m_r.
\end{align}
We could then expect that the asymptotic behaviour of the displacement square in diffusive regime would be the same as in non-relativistic case (\ref{displacement}). Indeed, we found that our Holographic calculation (\ref{dq1}), which is basically relativistic, showed that they are the same. 

However, the displacement square of our Holographic calculation in balistic regime gets multiplied by a factor of $\gamma^{-2}$ after substituting $m_0=m_r/\gamma$. Naively, we would expect that both the regimes changes in the same way. If we define an effective mass
\begin{align}
\label{eff mass}
 m_{eff}\equiv{m_0\over (1-w^2)^{3/2}}
\end{align}
such that an effective friction coefficient defined as
\begin{align}
 \gamma_{eff}=\gamma_0 (1-w^2)^{3/2}\to D={T\over \gamma_{eff} m_{eff}} \label{eff friction}
\end{align}
thus we obtain that both the regimes does not change their forms. Another reason that both regime should retain their forms is because we have computed the fluctuation in the rest frame of the string that co-rotating with the BTZ black hole. In this co-rotating frame, the string profile is a straight line that moving with a constant angular terminal velocity. So, locally the calculation is more or less similar to the case of non-rotating BTZ black hole as in~\cite{deBoer:2008gu} regardless of the effective geometry (\ref{eff geom}) is different from the non-rotating BTZ black hole. Hence the displacement square must have the similar form as in (\ref{displacement}) with observables, such as mass and friction coefficient, are defined in their relativistic form. Therefore we are tempted to identify the effective mass as an relativistic mass. An upshot of this we found that the relativistic mass of the external quark is multiplied by the Lorentz factor $\gamma$ to the power three instead of the usual $\gamma^{1}$, $m_r=\gamma^3 m_0$. This peculiar power of the Lorentz factor also appears in a study of the screening length in four dimensional Kerr-AdS black hole~\cite{Atmaja:2011ji} though the related observable is the mass of the black holes. 

As a resolution to this problem, we took an approach to analyze the Brownian motion through low frequency limit of the random-random force correlator. The calculation is more simple since we only have to look at the leading order in low frequency limit of the random-random force correlator and check if it satisfies equation (\ref{random 2 general})
. Another advantage of this approach is that the formula (\ref{random 2 general}) does not change its form in case of relativistic or, in general, under any defined effective mass, e.g. (\ref{eff mass}), in condition that we also redefine the friction coefficient to an effective friction coefficient, e.g. (\ref{eff friction}).

One could also argue that the peculiar power of Lorentz factor appears because our calculation is not fully relativistic since the Brownian motion we considered here is non-relativistic as described in Appendix \ref{Appendix 1}. Therefore one need to consider the relativistic Brownian motion to get the right asymptotic behaviour of the displacement square~\cite{Dunkel:2006,Dunkel:2009}. Further more, we have assumed that the Brownian motion or Langevin equation is linear in our calculation while the friction coefficient (\ref{rotation friction}) depends on angular velocity. For angular velocity dependent friction coefficient, one need to consider non-linear Brownian motion with the following non-linear Langevin equation~\cite{Lindner:2007}
\begin{align}
 m{du\over dt}=-\gamma(u)u+\psi(u)\xi(t),
\end{align}
where $\gamma(u)$ is the velocity dependent friction coefficient, $\psi(u)$ is noise function, and $\xi(t)$ is the noise. There has to be some relations between friction coefficient and the noise function for a given statistic of the noise by a general fluctuation-dissipation theorem~\cite{Stratonovich:1992}. One of those relations is derived by Dubkov, H\"{a}nggi, and Goychuk in~\cite{Dubkov:2009} such that the friction coefficient is constant only if the noise function is an additive noise source, $\psi(u)=1$, for white Gaussian noise, $\vev{\xi(t)\xi(t')}\sim\delta(t-t')$, and so it will reproduce the Einstein-Sutherland relation. However, the relativistic and/or non-linear Brownian motion are beyond our current work.

Nevertheless, it is interesting to study why the power of Lorentz factor behaves unusual for observable such as mass in the compact coordinate of rotating black holes. Another interesting thing is to check whether this behaviour still holds for higher dimensional rotating black holes and/or for charged rotating black holes. These will be explored more in our future works.

\section*{Acknowledgments}

 We would like to thank the organizer of CTPNP 2012,UGM-Yogyakarta, for the opportunity to present some of our work. 

\appendix

\section{Non-relativistic Brownian Motion}
\label{Appendix 1}
A main attention in the formulation of Brownian motion has always been about computation of the mean square value of the displacement of particle. The formula was first derived by Einstein in the case of free particle for late times~\cite{Einstein:1905}. A generalization for all times, in our notation, was given by Ornstein and F\"{u}rth independently~\cite{Ornstein:1917}:
\begin{align}
 s^2={2 k T \over \gamma_0^2 m}\left(\gamma_0 t-1+e^{-\gamma_0t}\right),
\end{align}
where $k$ is Boltzmann's constant which was set to be one in our Holographic Brownian motion. There are several ways to derived it and in this section we are going to use the momentum method following the derivation by Uhlenbeck and Ornstein~\cite{Uhlenbeck:1930zz}.

Before computing the displacement square, we are first calculating the mean (square) value of velocity $\vev{u}\left(\vev{u^2}\right)$. Rewriting the Langevin equation (\ref{BM}) without external force as follows:
\begin{align}
 {du\over dt}+\gamma_0 u= A(t),
\end{align}
with $A(t)=R(t)/m$, we find the solution
\begin{align}
\label{sol u}
 u=u_0 e^{-\gamma_0 t}+e^{-\gamma_0 t}\int_0^t e^{\gamma_0 \zeta}A(\zeta)d\zeta,
\end{align}
where $u_0$ is the velocity of Brownian particle at $t=0$. Its mean value is given by
\begin{align}
 \vev{u}^{u_0}&=u_0 e^{-\gamma_0 t},\label{u}\\
\vev{u^2}^{u_0}&=u_0^2 e^{-2\gamma_0 t}+e^{-2\gamma_0 t}\int_0^t\int_0^t e^{\gamma_0 (\zeta_0+\zeta_1)}\vev{A(\zeta_0)A(\zeta_1)}d\zeta_0 d\zeta_1.\label{u2}
\end{align}
Defining new variables $\zeta_0+\zeta_1=v$ and $\zeta_0-\zeta_1=w$, the integral part in (\ref{u2}) becomes
\begin{align}
 {1\over 2m^2}e^{-\gamma_0 t}\int_0^{2t} e^{\gamma_0 t} dv \int_{-\infty}^{+\infty}\rho(w)~ dw={\tilde{\rho}(0)\over 2\gamma_0 m^2}\left(1-e^{-2\gamma_0 t}\right).
\end{align}
Here,we have taken the second integral from $\int_{-t}^t dw$ to $\int_{-\infty}^{+\infty}dw$ under an assumption that $\rho(t)$ is considered as rapidly decreasing function. A requirement from equipartition theorem, at late times, the mean square of velocity must have
\begin{align}
 \lim_{t\to\infty}\vev{u^2}^{u_0}={k T\over m}
\end{align}
 and so we obtain a constant
\begin{align}
 \int_{-\infty}^{+\infty}\rho(w)~ dw=\tilde{\rho}(0)=2\gamma_0 m k T.
\end{align}

Integrating (\ref{sol u}) to get the position or displacement
\begin{align}
 s=x-x_0={u_0 \over \gamma_0}(1-e^{-\gamma_0 t})-{e^{-\gamma_0 t}\over \gamma_0}\int_0^t e^{\gamma_0\zeta}A(\zeta)d\zeta+{1\over\gamma_0}\int_0^t A(\zeta)d\zeta,
\end{align}
where $x_0$ is the initial position of the Brownian particle at $t=0$. The mean value of position is
\begin{align}
 \vev{s}^{u_0}&={u_0 \over \gamma_0}(1-e^{-\gamma_0 t}),\label{s}\\
 \vev{s^2}^{u_0}&={\tilde{\rho}(0)\over \gamma_0^2 m^2}t+{u_0^2\over \gamma_0^2}\left(1-e^{-\gamma_0 t}\right)^2+{\tilde{\rho}(0)\over 2\gamma_0^3 m^2}\left(-3+4e^{-\gamma_0 t}-e^{-2\gamma_0 t}\right). \label{s2}
\end{align}
Taking the average over initial velocity $u_0$, with $\overline{u_0}=0$ and $\overline{u_0^2}={kT\over m}={\tilde{\rho}(0)\over 2\gamma_0 m^2}$, the displacement becomes
\begin{align}
 \overline{\vev{s}}&=0,\\
 \overline{\vev{s^2}}&={\tilde{\rho}(0)\over \gamma_0^3 m^2}\left(\gamma_0 t -1+ e^{-\gamma_0 t}\right).
\end{align}
 Hence we obtain formula for the mean square displacement at all times and so its asymptotic time behaviour is given by (\ref{displacement}). In our Holographic calculation we use the natural unit $k=1$.

\end{document}